\documentclass[conference]{IEEEtran}
\IEEEoverridecommandlockouts
\usepackage{cite}
\usepackage{amsmath,amssymb,amsfonts}
\usepackage{algorithmic}
\usepackage{graphicx}
\usepackage{cuted}
\usepackage{textcomp}
\usepackage{url}
\usepackage{xcolor}
\usepackage{subfigure}
\usepackage{setspace}
\usepackage[ruled]{algorithm2e}
\def\BibTeX{{\rm B\kern-.05em{\sc i\kern-.025em b}\kern-.08em
    T\kern-.1667em\lower.7ex\hbox{E}\kern-.125emX}}
\usepackage{fancyhdr}
\usepackage[colorlinks, bookmarksopen, bookmarksnumbered, citecolor=blue, linkcolor=blue, urlcolor=blue]{hyperref}

\begin{document}

\title{Generalization Ability Analysis of Through-the-Wall Radar Human Activity Recognition}
\makeatletter
\newcommand{\linebreakand}{%
  \end{@IEEEauthorhalign}
  \hfill\mbox{}\par
  \mbox{}\hfill\begin{@IEEEauthorhalign}
}
\makeatother

\author{
  \IEEEauthorblockN{1\textsuperscript{st} Weicheng Gao}
  \IEEEauthorblockA{\textit{School of Information and Electronics} \\
    \textit{Beijing Institute of Technology (BIT)}\\
    Beijing, China \\
    JoeyBG@126.com}
  \and
  \IEEEauthorblockN{2\textsuperscript{nd} Xiaodong Qu*}
  \IEEEauthorblockA{\textit{School of Information and Electronics} \\
    \textit{Beijing Institute of Technology (BIT)}\\
    Beijing, China \\
    xdqu@bit.edu.cn}
  \and 
  \IEEEauthorblockN{3\textsuperscript{rd} Xiaopeng Yang}
  \IEEEauthorblockA{\textit{School of Information and Electronics} \\
    \textit{Beijing Institute of Technology (BIT)}\\
    Beijing, China \\
    xiaopengyang@bit.edu.cn}   
}

\maketitle
\thispagestyle{fancy}

\begin{abstract}
Through-the-Wall radar (TWR) human activity recognition (HAR) is a technology that uses low-frequency ultra-wideband (UWB) signal to detect and analyze indoor human motion. However, the high dependence of existing end-to-end recognition models on the distribution of TWR training data makes it difficult to achieve good generalization across different indoor testers. In this regard, the generalization ability of TWR HAR is analyzed in this paper. In detail, an end-to-end linear neural network method for TWR HAR and its generalization error bound are first discussed. Second, a micro-Doppler corner representation method and the change of the generalization error before and after dimension reduction are presented. The appropriateness of the theoretical generalization errors is proved through numerical simulations and experiments. The results demonstrate that feature dimension reduction is effective in allowing recognition models to generalize across different indoor testers.\par
\end{abstract}


\IEEEpeerreviewmaketitle

\section{Introduction}
\IEEEPARstart{T}{hrough-the-wall} radar (TWR), especially its applications on indoor human activity recognition (HAR), is mainly used in disaster rescue, anti-terrorism, and health monitoring, which attracts many teams to take effort mining cutting-edge algorithm designs \cite{TWR-Main1}. The micro-Doppler signature generated in the echoes can be used to achieve accurate motion feature extraction and activity recognition \cite{TWR-MCAE}.\par
Recent research results in the field of TWR HAR focused on the use of micro-Doppler signature to classify human activities, modeling different body parts of people behind walls, distinguishing, and predicting their intentions \cite{TWR-Main1}. These methods ranged from heuristic models \cite{SVM} to more sophisticated statistical learning techniques \cite{KNN, CRF}, including principal component analysis (PCA), independent component analysis (ICA), empirical modal decomposition (EMD), and Hilbert-Huang transform (HHT). Frontier research had focused on the use of deep convolutional neural networks (CNN) for single-stage feature extraction and activity recognition through large-scale data training \cite{GoogleNet VGG ResNet}. However, to solve the problem of high-precision extraction of fuzzy features from TWR imaging, the CNN-based recognition methods were improved by widening or deepening strategies, leading to a significant increase in spatial and temporal costs. Therefore, many lightweighting works were investigated \cite{Lightweight1, Lightweight2}.\par
\begin{figure*}
    \centering
    \includegraphics[width=\textwidth]{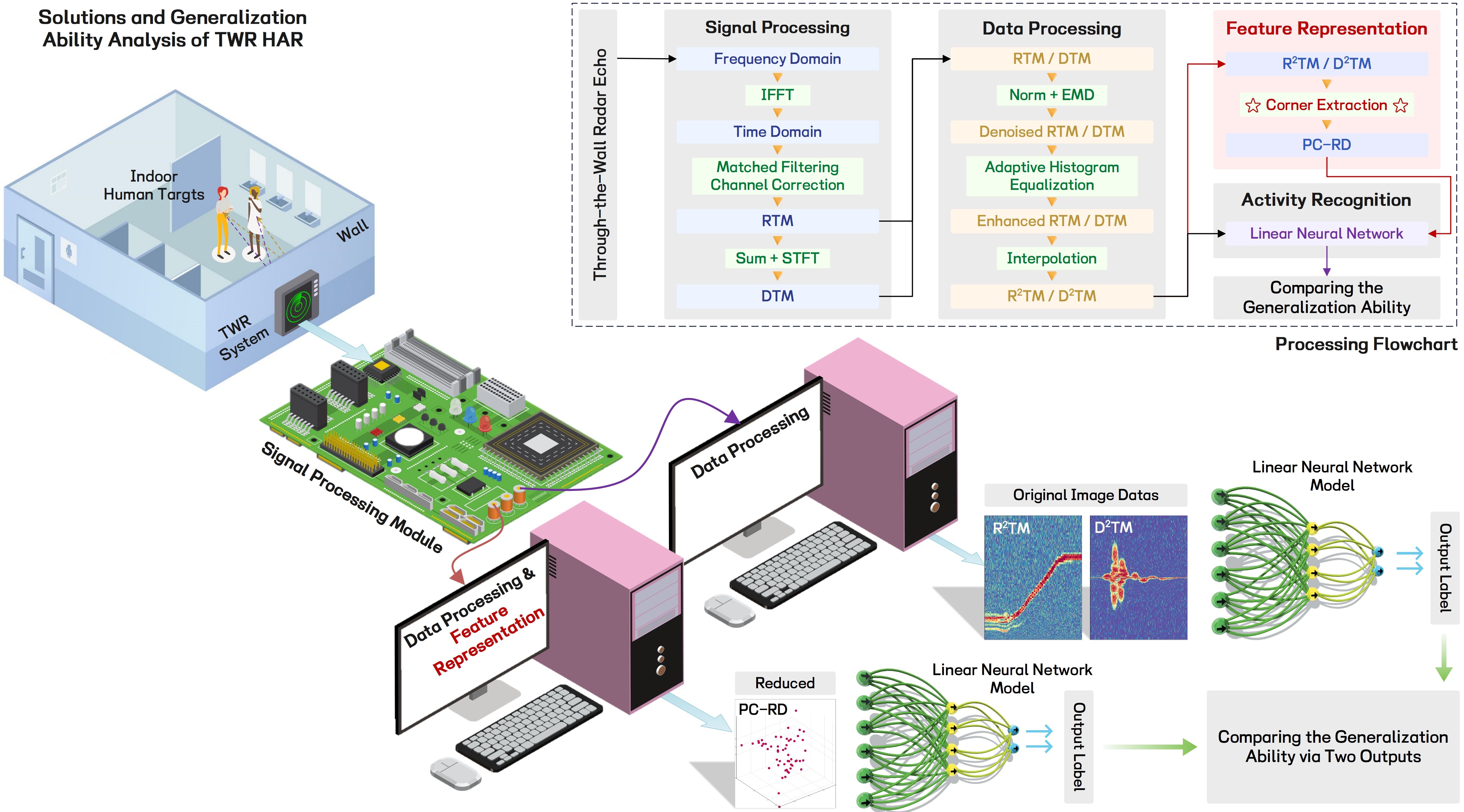}
    \caption{Solutions and generalization ability analysis for TWR HAR, where the feature representation is achieved using micro-Doppler corner detection. The signal / data processing modules and the overall structure of the recognition network are the same for both methods.}
    \label{Two Solutions}
    \vspace{-0.0cm}
\end{figure*}\par
\begin{figure*}
    \centering
    \includegraphics[width=\textwidth]{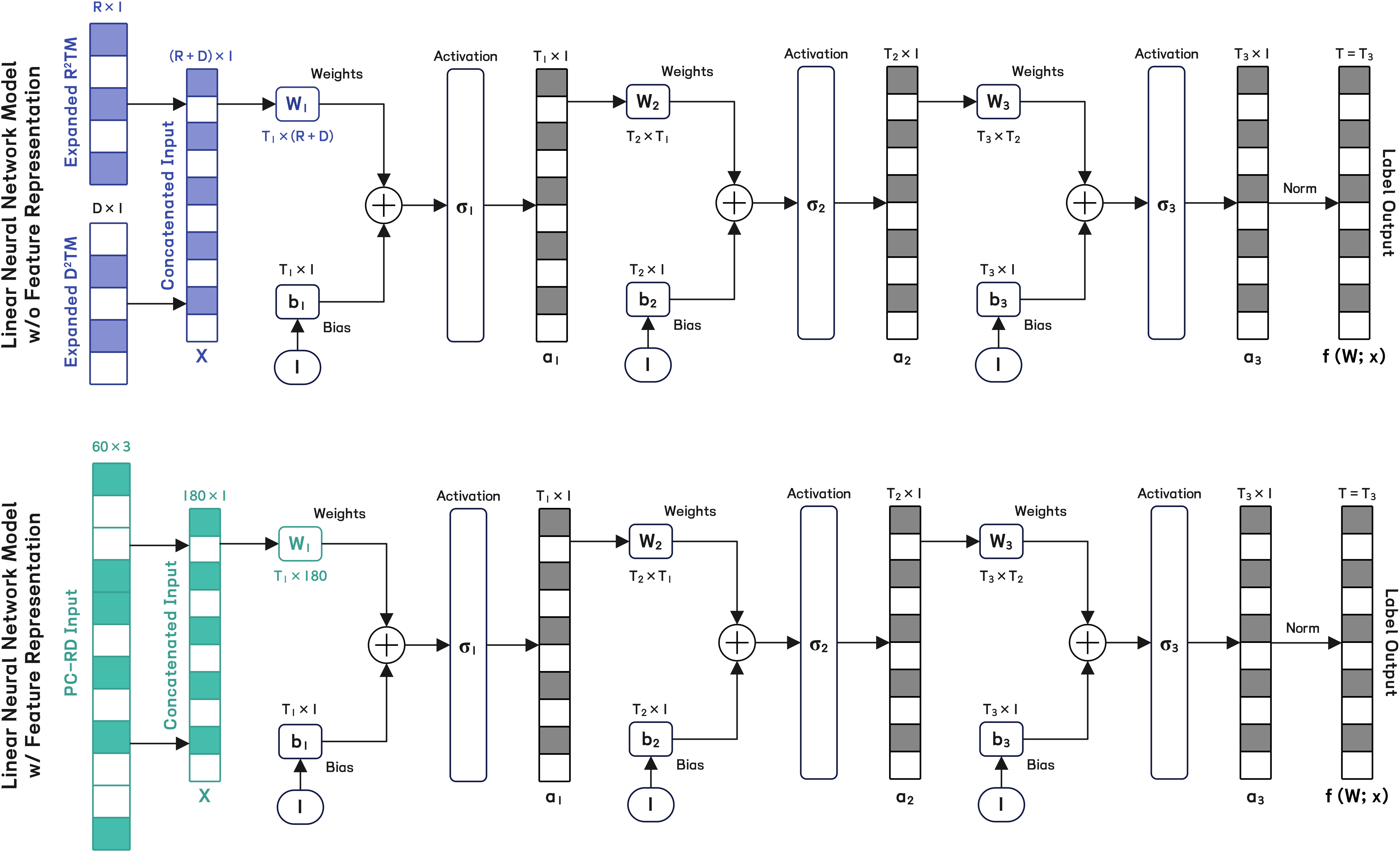}
    \caption{The design of the linear neural network models with and without feature representation.}
    \label{MLP}
    \vspace{-0.0cm}
\end{figure*}\par
Although existing methods address the issues of accuracy and resource cost, the models were highly dependent on the distribution of training data and difficult to generalize across various indoor testers. In our previous works, a series of generalized indoor HAR methods based on the downscaling of micro-Doppler corner feature were proposed \cite{Journal of Radars, Micro-Doppler Corner Detection}. However, none of them presented a rigorous discussion on how much feature dimension reduction could actually optimize the model generalization error in theory. Therefore, the generalization ability analysis of TWR HAR is presented in this paper. In detail, an end-to-end neural network approach for TWR HAR is proposed, and its micro-Doppler corner representation improvement is presented. In addition, the analysis of the generalization error bounds of both methods are given, respectively, and the enhancement of model generalization ability by dimension reduction is theoretically demonstrated. Moreover, the reasonableness of the theoretical analysis is verified through numerical simulations and experiments.\par

\section{Feature Reduction Based on Micro-Doppler Corner Representation}
As shown in Fig. \ref{Two Solutions}, the generalization ability of TWR HAR is compared between two different features. The first is the \textbf{TWR range and Doppler profiles} directly generated by the signal and data processing modules. The second feature is the \textbf{micro-Doppler corner representation}, which is obtained by generating TWR range and Doppler profiles using the same signal and data processing modules and extracting the key points' coordinates on them using corner detection algorithm. Both features are fed into linear neural networks for recognition, and the generalization performance of them is compared by the training effect.\par
In signal processing module, the raw TWR echo in the frequency domain is first captured. Time-domain raw echo is then generated using the inverse Fourier transform (IFFT). Then, the extraction of the baseband time-domain echo signal is achieved by local oscillation mixing and low-pass filtering. The baseband echoes over a slow period of time are concatenated into a matrix, which is called range-time map (RTM). The RTM is summed along the range axis and the short-time Fourier transform (STFT) is operated to obtain the Doppler-time map (DTM) \cite{Lightweight2}.\par
In data processing module, both RTM and DTM are input. After normalization (Norm) and EMD processing, the noise on both the images is suppressed to a certain degree. Adaptive histogram equalization algorithm is then applied to achieve contrast enhancement of micro-Doppler signature. Finally, the range axis of the enhanced RTM, and the Doppler axis of the enhanced DTM, are interpolated to stretch its units from linear to square. The resulting $\mathbf{R^2TM}$ and $\mathbf{D^2TM}$ are served for subsequent feature extraction and activity recognition \cite{Journal of Radars}.\par
For the traditional method, the $\mathbf{R^2TM}$ and $\mathbf{D^2TM}$ are directly fed into the network. Here, a feature reduction method based on micro-Doppler corner representation is proposed. In detail, both $\mathbf{R^2TM}$ and $\mathbf{D^2TM}$ are converted into a two-dimensional micro-Doppler corner point cloud representation using Gaussian Differential Convolution (DoG). $30$ corner points are detected on both images and concatenated into a three-dimensional point cloud. Information in default dimension is found by non-maximum suppression (NMS) on the other map. The final obtained micro-Doppler corner representation $\mathbf{PC-RD}$ is a $60 \times 3$ vector \cite{Journal of Radars}.\par
The activity recognition is finally achieved by linear neural networks \cite{MLP GEB}. As shown in Fig. \ref{MLP}, in traditional method, the $\mathbf{R^2TM}$ and $\mathbf{D^2TM}$ are first flattened into vectors and concatenated. Then, the concatenated vector is processed through three separate fully-connected layers with activation functions and Norm to achieve feature-to-label mapping. In feature representation method, the $\mathbf{PC-RD}$ is also fed into the linear neural network model after being stretched into a $180 \times 1$ sequence. The structure and parameters of the other layers are kept consistent with the network in traditional method.\par

\section{Generalization Performance Analysis}
As shown in Fig. \ref{MLP}, assuming the concatenated input dataset to the model is $x$. Then the true loss to the model is:

\vspace{-0.2cm}
\begin{equation}
L_\mathrm{true}(f)=\mathbb{E}_{z \thicksim P_x}[L(f,z)],
\end{equation}
where $P_x$ is the distribution of data $x$. $L$ denotes the loss of model $f$ on a single data $z$ satisfying distribution $P_x$, where:

\vspace{-0.2cm}
\begin{equation}
f(W;x)=\sigma_3(W_3(\sigma_{2}W_{2}(\sigma_{1}(W_1 x)))) \overset{\mathrm{Def}}{=}\circ_{i=1}^3\sigma_iW_i(x),
\end{equation}
is the transfer function of the model. $W_i,~i=1,2,3$ are the weights and $\sigma_i,~i=1,2,3$ are the activation functions in each layer. Similarly, the training loss of the model is:

\vspace{-0.2cm}
\begin{equation}
\begin{aligned}
L_\mathrm{train}(f;S)=\frac{1}{|S|}\sum_{\mathrm{All}~i} L(f,z_i)
\end{aligned},
\end{equation}
where $z_i$ is the $i^\mathrm{th}$ sample in training data $S$. The generalization error of the model is defined as the difference between the true loss and the training loss \cite{MLP GEB}:

\vspace{-0.2cm}
\begin{equation}
\mathrm{GE} = L_\mathrm{true}(f) - L_\mathrm{train}(f;S).
\end{equation}\par
In order to be able to quantify the bounds on the generalization error, the activation function in the model can all be assumed as Lipschitz continuous with the constant of $C$, and the loss function $L$ is assumed as Lipschitz continuous with the constant of $L^{\prime}$. Define the maximum eigenvalue of $W_i$ as $\lambda_i$. According to the triangular inequality, for any real vector $x,y$ with the same rows of $W_i$:

\vspace{-0.2cm}
\begin{equation}
\begin{aligned}
\left\|W_i(x)-W_i(y)\right\|_2=\left\|W_i(x-y)\right\| &\leq \left\|\lambda_i(x-y)\right\|_2\\&=\lambda_i \left\|x-y\right\|_2
\end{aligned}.
\end{equation}\par
Thus, define $y=x+dx$ in the equation above, the sensitivity of the model to the data can be measured as:

\vspace{-0.2cm}
\begin{equation}
\begin{aligned}
&\left\|f(W;x+d x)-f(W;x)\right\|_2\\
&\! =\left\| \circ_{i=1}^3\sigma_i W_i(x+d x)-\circ_{i=1}^3\sigma_i W_i(x) \right\|_2\\
&\! \leq C\lambda_3\left\|\circ_{i=1}^{2}\sigma_i W_i(x+\delta x)-\circ_{i=1}^{2}\sigma_i W_i(x)\right\|_2\\
&\! \leq C^2\lambda_3\lambda_2\left\|\sigma_1 W_1(x+\delta x)-\sigma_1 W_1(x)\right\|_2\\
&\! \leq C^3 \prod_{i=1}^3 \lambda_i \prod_{i=1}^3 \left\|x+d x-x\right\|_2 \leq C^3 \prod_{i=1}^3 \lambda_i \left\|dx\right\|_2
\end{aligned}.
\label{Data Sensitivity}
\end{equation}\par
According to \cite{MLP GEB}, the sensitivity of the model is:

\vspace{-0.2cm}
\begin{equation}
\begin{aligned}
&||f(W+d W;x)-f(W;x)||_2^2\\
&\! \leq \left(2\frac{\beta}{\alpha}\mathrm{\omega}\kappa^2\right)^3 \bigg(\prod_{i=1}^3\left(1+\frac{||d W_i||_F^2}{||W_i||_F^2}\right)-1\bigg)||f(W;x)||_2^2
\end{aligned},
\label{Weights Sensitivity}
\end{equation}
where $\omega$ is the maximum width of the network, $\kappa$ is the condition number bound of $W_i$. $||\cdot||_F$ denotes the Frobenius norm. $\alpha,\beta$ are the lower and upper Lipschitz bounds of the activations. Because the data used is not simply Gaussian distributed, here the condition number is assumed as finite.\par
In the finite hypothesis space of the HAR task, the mathematical model of the generalization error bound for the training model can be established as the following inequality:

\vspace{-0.2cm}
\begin{equation}
\mathbb{E}_S[|L(f_S,z_i)-L(f_{S/z_i},z_i)|]\leq\beta 
\end{equation}
where $z_i\in S,~i=1,2,\ldots,M,~M\in \mathbb{Z^{+}}$ is the dataset. $S/z_i$ denotes the remaining dataset without $z_i$, $S/z_i \cup z_i = S$. The model obtained at the end of training is $f_S$. Define the model obtained at the end of training with $S/z_i$ is $f_{S/z_i}$. Thus:

\vspace{-0.2cm}
\begin{equation}
\begin{aligned}
& L(f_S,S)-L(f_{S/z_i},S/z_i)\\
&\! \leq L^{\prime} |f_S(S)-f_{S/z_i}(S/z_i)|  \\
&\! =L^{\prime} |f(W+d W;x+d x)-f(W;x) \\
&\! =L^{\prime} |f(W+d W;x+d x)-f(W+d W;x)\\
&\! \qquad +f(W+d W;x)-f(W;x)| \\
&\! \leq L^{\prime} (|f(W+d W;x+d x)-f(W+d W;x)|\\
&\! \qquad +|f(W+d W;x)-f(W;x)|)
\end{aligned}
\end{equation}
According to \cite{MLP GEB}, the sensitivity on both data and weights of the model can be denoted as:

\vspace{-0.2cm}
\begin{equation}
\begin{aligned}
&|f(W+d W;x+d x)-f(W+d W;x)|\\
&\! \leq\sqrt{h}\bigg(C^3 \prod_{i=1}^3 \lambda_i\bigg)||d x||_2
\end{aligned},
\label{Data and Weights Sensitivity}
\end{equation}
where $h$ is the number of activity labels.\par
According to Eq. (\ref{Weights Sensitivity}):

\vspace{-0.2cm}
\begin{equation}
\begin{aligned}
&|f(W+d W;x)-f(W;x)|\\
&\! \leq\sqrt{h}\sqrt{Q{\left(\prod_{i=1}^3\left(1+\frac{||\delta W_i||_F^2}{||W_i||_F^2}\right)-1\right)}||f(W;x)||_2^2}    
\end{aligned},
\end{equation}
where $Q=\left(2\frac{\beta}{\alpha}\mathrm{\omega}\kappa^2\right)^3$. Since both the input vector and the output label sequence are normalized:

\vspace{-0.2cm}
\begin{equation}
\begin{aligned}
&||d x||_2=1,\quad ||f(W;x)||_2^2=1\\
&|L(f_S,S)-L(f_{S/z_i},S/z_i)|\leq L^{\prime} \sqrt{h} \cdot \gamma_{S,S/z_i}
\end{aligned},
\end{equation}
where $\gamma_{S,S/z_i}= \gamma_{S,S/z_i,1}+\gamma_{S,S/z_i,2}$ yields:

\vspace{-0.2cm}
\begin{equation}
\begin{aligned}
\gamma_{S,S/z_i,1}&=C^3\prod_{i=1}^3\lambda_i\\
\gamma_{S,S/z_i,2}&=\sqrt{Q{\left(\prod_{i=1}^3\left(1+\frac{||d W_i||_F^2}{||W_i||_F^2}\right)-1\right)}}
\end{aligned}.
\end{equation}\par
Conventional training methods used for linear neural networks are gradient descent and its improvements \cite{MLP GEB}. For normalized inputs and outputs, in each training round, the change of weight matrices always meets:

\vspace{-0.2cm}
\begin{equation}
||d W_i ||_F^2\leq 1,
\end{equation}
for all $i$. Define the network is totally trained in $N$ rounds. Thus:

\vspace{-0.2cm}
\begin{equation}
\begin{aligned}
\gamma_{S,S/z_i,1}&=C^3\prod_{i=1}^3\lambda_i\\
\gamma_{S,S/z_i,2}&=\sqrt{Qh\left(\prod_{i=1}^3\left(1+\frac{N}{||W_i||_F^2}\right)-1\right)}
\end{aligned},
\end{equation}
which means that:

\vspace{-0.2cm}
\begin{equation}
\begin{aligned}
&|L(f_S,S)-L(f_{S/z_i},S/z_i)|\\
&\! \leq L^{\prime} \sqrt{h} \left(C^3\prod_{i=1}^3\lambda_i+ \sqrt{Qh\left(\prod_{i=1}^3\left(1+\frac{N}{||W_i||_F^2}\right)-1\right)}\right)
\end{aligned}.
\label{Loss Error Inequation}
\end{equation}\par
According to \cite{MLP GEB} and Eq. (\ref{Loss Error Inequation}), for probability greater than or equal to $1-\delta,~\delta\in \mathbb{R}^{(0,1)}$, the difference between the true loss and the training loss can be denoted in Eq. (\ref{Error Bound Inequation}), where $B$ is the bound of the loss function and $\mathrm{GEB}$ is the desired generalization error bound.\par
The process remains consistent for generalization error analysis of the model in “Method No.2”. Define the generalization error bound of the improved method is $\mathrm{GEB}^{I}$, the maximum width of the network is $\omega^{I}$, the weight matrices are $W^{I}_i,~i=1,2,3$ with their maximum eigenvalues $\lambda^{I}_i,~i=1,2,3$ and condition number bound $\kappa^{I}$. Eq. (\ref{GEBI}) presents the desired generalization error bound after dimension reduced improvement based on micro-Doppler corner representation.\par
\begin{figure*}
    \centering
    \includegraphics[width=\textwidth]{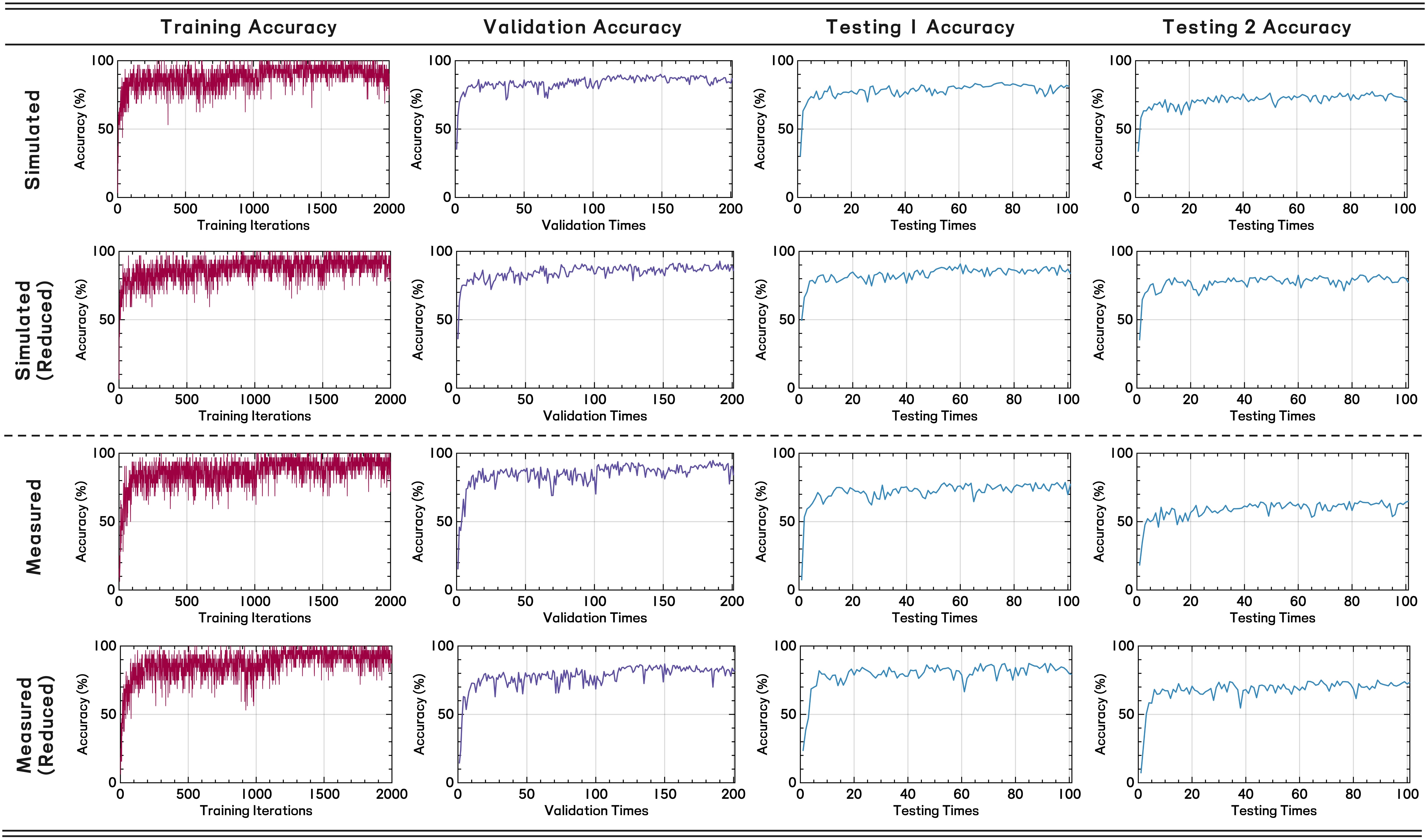}
    \caption{Plot of accuracy changing during model training. “Reduced” denotes “Feature reduction based on micro-Doppler corner representation”.}
    \label{Acc Plots}
    \vspace{-0.0cm}
\end{figure*}\par
\begin{figure*}
    \centering
    \includegraphics[width=\textwidth]{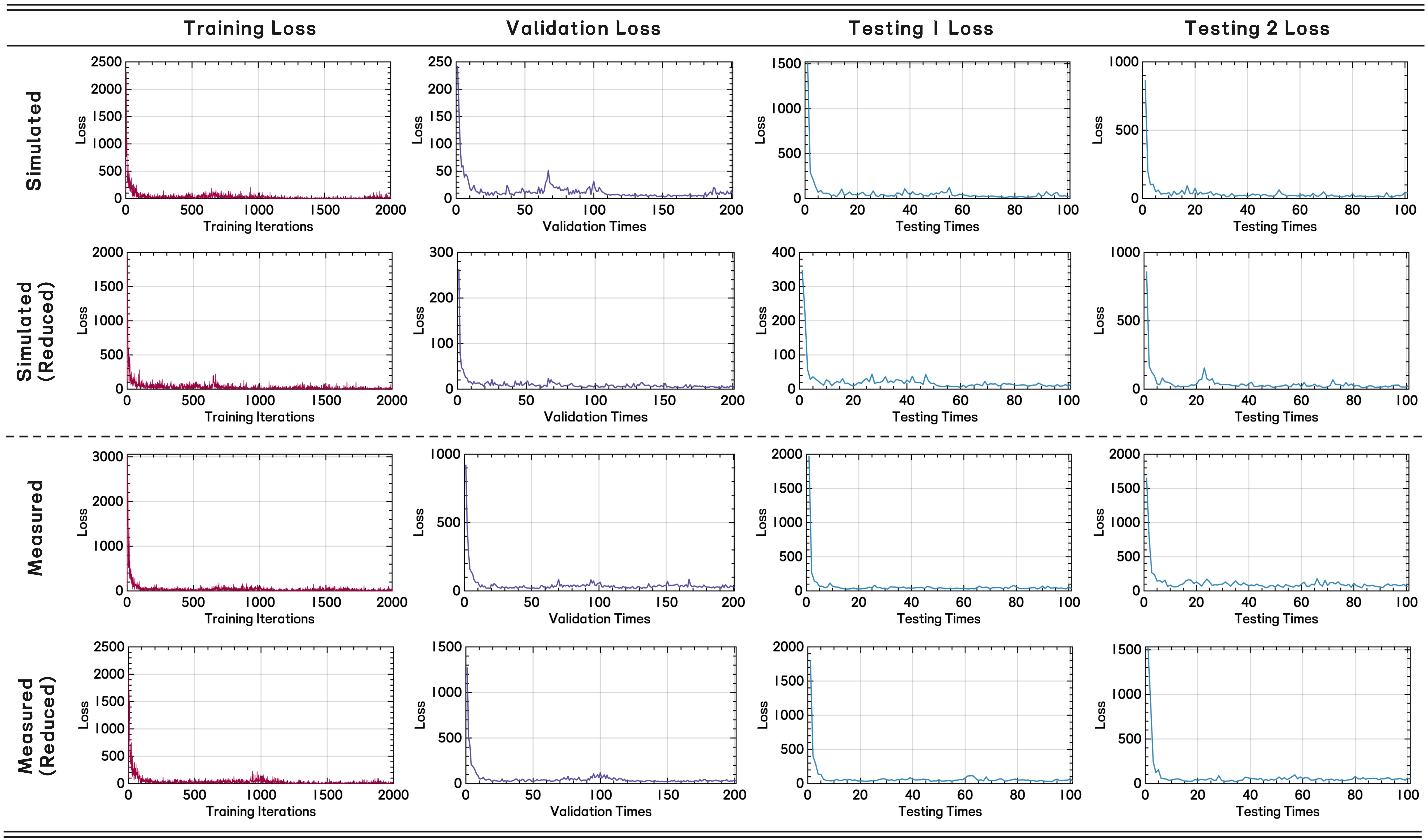}
    \caption{Plot of loss changing during model training. “Reduced” denotes “Feature reduction based on micro-Doppler corner representation”.}
    \label{Loss Plots}
    \vspace{-0.0cm}
\end{figure*}\par
To prove the generalization error bound decreases after dimension reduction, all that's required is to prove that $\frac{\mathrm{GEB}^{I}}{\mathrm{GEB}}<1$. Comparing Eq. (\ref{Error Bound Inequation}) and (\ref{GEBI}), both have the same denominator inside the root sign as well as the constant term in the numerator, thus the problem is equivalent to Eq. (\ref{Proof1}). The eigenvalues of both the left and right sides are then relaxed, and the necessary insufficiency condition for the proof can be rewritten as Eq. (\ref{Proof2}). In the proposed model, $\omega^{I}=180$, but $\omega=R+D$, where $R$ and $D$ are defined in Fig. \ref{MLP}. Thus, Eq. (\ref{Proof2}) can be then relaxed to $\sqrt{180^3}\ll \sqrt{(R+D)^3}-1$. Empirically, the value of $R+D$ is usually close to or even larger than $10^4$ \cite{Lightweight2}. Thus, $\sqrt{180^3}\ll \sqrt{(R+D)^3}-1$ holds. The above analysis theoretically demonstrates that the feature dimension reduction method based on micro-Doppler corner representation does improve the generalization ability of the TWR HAR model.\par
\begin{strip}
\hrulefill
\begin{equation}
\begin{aligned}
\mathrm{GE}=L_\mathrm{true}(f)-L_\mathrm{train}(f;S)\leq \mathrm{GEB}=\sqrt{\frac{B^2+12BML^{\prime}\sqrt{h}\left(C^3\prod_{i=1}^3\lambda_i+\sqrt{\left(2\frac{\beta}{\alpha}\mathrm{\omega}\kappa^2\right)^3 \left(\prod_{i=1}^3\left(1+\frac{N}{||W_i||_F^2}\right)-1\right)}\right)}{2M\delta}}
\end{aligned},
\label{Error Bound Inequation}
\end{equation}
\begin{equation}
\mathrm{GEB}^{I}=\sqrt{\frac{B^2+12BML^{\prime}\sqrt{h}\left(C^3\prod_{i=1}^3\lambda^{I}_i+\sqrt{\left(2\frac{\beta}{\alpha}\mathrm{\omega}^{I}{\kappa^{I}}^2\right)^3 \left(\prod_{i=1}^3\left(1+\frac{N}{||W^{I}_i||_F^2}\right)-1\right)}\right)}{2M\delta}}
\label{GEBI}
\end{equation}
\begin{equation}
C^3\prod_{i=1}^3\lambda^{I}_i+\sqrt{\left(2\frac{\beta}{\alpha}\mathrm{\omega}^{I}{\kappa^{I}}^2\right)^3 \left(\prod_{i=1}^3\left(1+\frac{N}{||W^{I}_i||_F^2}\right)-1\right)} \leq C^3\prod_{i=1}^3\lambda_i+\sqrt{\left(2\frac{\beta}{\alpha}\mathrm{\omega}\kappa^2\right)^3 \left(\prod_{i=1}^3\left(1+\frac{N}{||W_i||_F^2}\right)-1\right)}
\label{Proof1}
\end{equation}
\begin{equation}
\Leftarrow \sqrt{\left(2\frac{\beta}{\alpha}\mathrm{\omega}^{I}{\kappa^{I}}^2\right)^3 \left(\prod_{i=1}^3\left(1+\frac{N}{||W^{I}_i||_F^2}\right)-1\right)} \leq \sqrt{\left(2\frac{\beta}{\alpha}\mathrm{\omega}\kappa^2\right)^3 \left(\prod_{i=1}^3\left(1+\frac{N}{||W_i||_F^2}\right)-1\right)}-C^3
\label{Proof2}
\end{equation}
\vfill
\end{strip}\par

\section{Numerical Simulations and Experiments}
\subsection{System and Parameters}
In this paper, the generalization performance of the theoretically analyzed TWR HAR models is verified using both simulated and measured datasets. The simulated dataset is achieved using UWB signal transmitting in the frequency range of $0.5\sim 2.5\mathrm{~GHz}$. The wall parameters are set to be $12\mathrm{~cm}$ thick with a relative dielectric constant of $6$. Human motion information is implemented by an open source motion capture dataset from University College London (UCL) \cite{UCL}. The radar and wall environment used for the measured dataset is consistent with the simulation scenario. The captured data is fed into the signal and data processing module to generate $\mathbf{R^2TM}$, $\mathbf{D^2TM}$, and $\mathbf{PC-RD}$, all with $3200$ sets for training (“Training”: $1.8~m$ height human), $800$ sets for validation (“Validation”: $1.8~m$ height human), and $400$ sets for each tester (“Testing $1$”: $1.7~m$ height and “Testing $2$”: $1.6~m$ height human). Activity labels include: Empty; Punching; Kicking; Grabbing; Sitting Down; Standing Up; Rotating; Walking; Sitting to Walking; Walking to Sitting; Falling to Walking; Walking to Falling. Both network models pass $20$ epochs of training with a batch size of $32$ and a validation and testing frequency of $10$ batches at a time. Initial learning rate is $0.00147$. Parameters $R$ and $D$ in Fig. \ref{MLP} are both $4096$.\par

\subsection{Verification of Generalization Ability}
As shown in Fig. \ref{Acc Plots}, the models corresponding to both methods converge after no more than $20$ epochs of training. Comparing the four plots in the first line, which denote the results of simulated dataset under traditional method, the end-epoch accuracy of validation and testing is decreased compared to the end-epoch accuracy of training. The differences in accuracy between the validation set and the test sets are $6.00\%$ and $16.50\%$. The end-epoch loss of validation and testing is increased compared to the end-epoch loss of training. Comparing the four plots in the second line, which denote the results of simulated dataset under feature reduction based on micro-Doppler corner representation, the differences in accuracy between the validation set and the testing sets are both reduced to $2.13\%$ and $8.63\%$, respectively. Comparing the four plots in the third line, which denote the results of measured dataset under traditional method, the differences in accuracy between the validation set and the testing sets are $6.13\%$ and $18.88\%$, respectively. Comparing the four plots in the fourth line, which denote the results of measured dataset under feature reduction based on micro-Doppler corner representation, the differences in accuracy between the validation set and the testing sets are reduced to $3.25\%$ and $9.00\%$, respectively. Comparison of the convergence effects of the loss functions in Fig. \ref{Loss Plots} leads to exactly the same patterns. The above results prove that the generalization ability of the model is improved after the micro-Doppler corner representation based feature reduction. The enhancement level is also close to the theoretically derived generalization error results.\par

\section{Conclusion}
In order to address the problem that existing TWR HAR models have not generalized well across different indoor human testers, this paper has provided a theoretical analysis of TWR HAR's generalization performance. In detail, the generalization error bound of an end-to-end linear neural network technique has been calculated for TWR HAR. Also, the generalization error improvement before and after micro-Doppler corner representation based dimension reduction has been analyzed. Numerical simulated and measured experiments have been carried out to prove that the generalization ability analysis is valid. The results have shown that feature dimension reduction has worked well to enable recognition models to generalize across different indoor testers.\par

\section*{Acknowledgement}
This work was supported in part by the National Natural Science Foundation of China under Grant 62101042; and in part by Beijing Institute of Technology Research Fund Program for Young Scholars under Grant XSQD-202205005. (Corresponding author: Xiaodong Qu.)\par



\begin{thebibliography}{00}
\bibitem{TWR-Main1}N. Maaref, P. Millot, C. Pichot and O. Picon, “A Study of UWB FM-CW Radar for the Detection of Human Beings in Motion Inside a Building,” \emph{IEEE Trans. Geosci. Remote Sens.}, vol. 47, no. 5, pp. 1297-1300, May 2009.
\bibitem{TWR-MCAE}W. Gao, X. Yang, X. Qu and T. Lan, “TWR-MCAE: A Data Augmentation Method for Through-the-Wall Radar Human Motion Recognition,” \emph{IEEE Trans. Geosci. Remote Sens.}, vol. 60, pp. 1-17, 2022, Art no. 5118617.
\bibitem{SVM}L. Mohan, J. Pant, P. Suyal and A. Kumar, “Support Vector Machine Accuracy Improvement with Classification,” in \emph{Proc. Intern. Conf. Comput. Intell. Commun. Netw.}, 2020, pp. 477-481.
\bibitem{KNN}R. Yu, Y. Du, J. Li, A. Napolitano, and J. Le Kernec, “Radar‐based human activity recognition using denoising techniques to enhance classification accuracy,” \emph{IET Radar Sonar Navig.}, vol. 18, no. 2, pp. 277–293, Nov. 2023.
\bibitem{CRF}X. Qu, W. Gao, H. Meng, Y. Zhao and X. Yang, “Indoor Human Behavior Recognition Method Based on Wavelet Scattering Network and Conditional Random Field Model,” \emph{IEEE Trans. Geosci. Remote Sens.}, vol. 61, pp. 1-15, 2023, Art no. 5104815.
\bibitem{GoogleNet VGG ResNet}F. J. Abdu, Y. Zhang and Z. Deng, “Activity Classification Based on Feature Fusion of FMCW Radar Human Motion Micro-Doppler Signatures,” \emph{Sens. J.}, vol. 22, no. 9, pp. 8648-8662, 1 May1, 2022.
\bibitem{Lightweight1}X. Yang, W. Gao, X. Qu, P. Yin, H. Meng, and A. E. Fathy, “A Lightweight Multiscale Neural Network for Indoor Human Activity Recognition Based on Macro and Micro-Doppler Features,” \emph{IEEE Intern. Things J.}, vol. 10, no. 24, pp. 21836–21854, Dec. 2023.
\bibitem{Lightweight2}X. Wang, S. Guo, J. Chen, P. Chen, and G. Cui, “GCN-Enhanced Multidomain Fusion Network for Through-Wall Human Activity Recognition,” \emph{IEEE Geosci. Remote Sens. Lett.}, vol. 19, pp. 1–5, 2022.
\bibitem{Journal of Radars}X. Yang, W. Gao, X. Qu, “Human Anomalous Gait Termination Recognition via Through-the-wall Radar Based on Micro-Doppler Corner Features and Non-Local Mechanism,” \emph{J. Radars}, vol. 13, no. 1, pp. 68-86, 2024.
\bibitem{Micro-Doppler Corner Detection}W. Gao, X. Qu, H. Meng, X. Sun and X. Yang, “Adaptive Micro-Doppler Corner Feature Extraction Method Based on Difference of Gaussian Filter and Deformable Convolution,” \emph{IEEE Signal Process. Lett.}, vol. 31, pp. 860-864, 2024.
\bibitem{MLP GEB}X. Lan, X. Guo and K. E. Barner, “PAC-Bayesian Generalization Bounds for MultiLayer Perceptrons,” \emph{Arxiv}, 16-Jun-2020. [Online]. Available: https://arxiv.org/abs/2006.08888.
\bibitem{UCL}S. Vishwakarma, W. Li, C. Tang, K. Woodbridge, R. Adve and K. Chetty, “SimHumalator: An Open-Source End-to-End Radar Simulator for Human Activity Recognition,” \emph{IEEE Aeros. Electron. Syst. Mag.}, vol. 37, no. 3, pp. 6-22, 1 March 2022.
\end{thebibliography}
\end{document}